# Lyapunov exponents from geodesic spread in configuration space

Monica Cerruti-Sola[†]

*Osservatorio Astrofisico di Arcetri, Largo E. Fermi 5, 50125 Firenze, Italy*

Roberto Franzosi[*]

*Dipartimento di Fisica, Università di Firenze, Largo E. Fermi 2, 50125 Firenze, Italy*

Marco Pettini[‡]

*Osservatorio Astrofisico di Arcetri, Largo E. Fermi 5, 50125 Firenze, Italy*

(June 28, 2018)

## Abstract

The exact form of the Jacobi – Levi-Civita (JLC) equation for geodesic spread is here explicitly worked out at arbitrary dimension for the configuration space manifold $M_E = \{q \in \mathbb{R}^N | V(q) < E\}$ of a standard Hamiltonian system, equipped with the Jacobi (or kinetic energy) metric $g_J$. As the Hamiltonian flow corresponds to a geodesic flow on $(M_E, g_J)$, the JLC equation can be used to study the degree of instability of the Hamiltonian flow. It is found that the solutions of the JLC equation are closely resembling the solutions of the standard tangent dynamics equation which is used to compute Lyapunov exponents. Therefore the instability exponents obtained through the JLC equation are in perfect quantitative agreement with usual Lyapunov exponents. This work completes a previous investigation that was limited only to two-degrees of freedom systems.

Typeset using REVTEX

In recent papers [1,2] we have investigated the dynamical stability properties of two-degrees of freedom Hamiltonians ($N = 2$) within the framework of a geometric formulation of dynamics that makes use of Riemannian geometry. At $N = 2$ the phase space structure of a system can be investigated in great detail. In fact the use of Poincaré surfaces of section makes possible to identify the initial conditions that originate regular and chaotic motions in the system, so that this qualitative description as well as the measurement of chaos by Lyapunov exponents can be thoroughly compared with the outcome of the Riemannian based approach. However, the $N = 2$ case is a very special case, at least from the geometric point of view; in fact there is only one curvature function that - at each point - plays the role of scalar curvature, Ricci curvature and sectional curvature. Therefore, *in absence of any rigorous result* to extend at arbitrary $N$ the validity of what we found at $N = 2$, we have explicitly studied the large $N$ case and the results are given in the present paper. There is also another motivation for the present work. We have recently exploited the Riemannian geometrization of newtonian dynamics to *analytically* compute the largest Lyapunov exponents in large-$N$ Hamiltonian systems [3–5], and, despite some necessary approximation, the analytic results are in strikingly good agreement with the numerical results. However, while applying this theory to lattice-$\varphi^4$ models [6] we have encountered some difficulties that are now demanding adequate improvements. For the sake of simplicity, all the analytic computations were done in an enlarged configuration space-time endowed with Eisenhart metric (see below). In this framework, the mentioned improvements can hardly be imagined and a richer geometric structure, as is the case of $(M_E, g_J)$, is needed (we shall better explain why in the sequel). Therefore it is of primary importance to check whether the JLC equation on $(M_E, g_J)$ fully accounts for the degree of chaoticity of the dynamics at arbitrary $N$. In principle this might not be the case: the JLC equation only describes *local* instability, whereas chaos *could* crucially depend upon some *global* property of phase space. As a simple example, let us think of the Bunimovich stadium (a portion of the plane, bounded by two half-circles joined by two parallel lines, where a free particle bounces), where the shape of the boundary, being responsible for the mismatch between focusing and defocusing of trajectories, makes the system chaotic. In the case of Hamiltonian flows at $N = 2$, something similar happens when a trajectory approaches the condition $V(q) = E$: the curvature function becomes very large because it contains powers of the quantity $[E - V(q)]^{-1}$ and, correspondigly, the configuration space trajectories look as if they were reflected by the $V(q) = E$ boundary. At large $N$ such a stadium-like effect is *no longer* present and $[E - V(q)]$ fluctuates around an average value with a negligible probability of getting close to zero, therefore "global" effects – if any – should work in a subtler way.

We consider those systems that are decribed by the Lagrangian function (all the indexes run from 1 to $N = dim M_E$)

$$L(q, \dot{q}) = \frac{1}{2} a_{ik}(q) \dot{q}^i \dot{q}^k - V(q) , \tag{1}$$

where $a_{ik}$ is the kinetic energy tensor

$$a_{ik} \dot{q}^i \dot{q}^k = 2(E - V) = 2W . \tag{2}$$

Maupertuis' least action principle

$$\delta \int_\gamma 2W \, dt = \delta \int_\gamma \{2[E - V(q)] a_{ik} dq^i dq^k\}^{1/2} \equiv \delta \int_\gamma ds = 0 \tag{3}$$

variationally defines the natural motions among all the isoenergetic asynchronous paths $\gamma$ joining two fixed endpoints. Hence the arc-length of configuration space is expressed by $ds^2 = 2[E - V(q)] a_{ik} dq^i dq^k$ whence $g_{ik} = (E - V(q)) a_{ik}$. In local coordinates the geodesic equations on a Riemannian manifold are solutions of the equations

$$\frac{d^2 q^i}{ds^2} + \Gamma^i_{jk} \frac{dq^j}{ds} \frac{dq^k}{ds} = 0 \tag{4}$$

where $s$ is the proper time and $\Gamma^i_{jk}$ are the Christoffel coefficients of the Levi-Civita connection associated with $g_{ik}$. By direct computation, using $g_{ik} = (E - V(q)) \delta_{ik}$, $\Gamma^i_{jk} = \frac{1}{2W} \delta^{im} (\partial_j W \delta_{km} + \partial_k W \delta_{mj} - \partial_m W \delta_{jk})$ and $ds^2 = 2W^2 dt^2$, it can be easily verified that the geodesic equations yield

$$\frac{d^2 q^i}{dt^2} = -\frac{\partial V}{\partial q^i} \quad i = 1, \ldots, N \tag{5}$$

i.e. Newton's equations derived from the Lagrangian (1). These equations of motion can be also seen as geodesics of other manifolds [7] besides $(M_E, g_J)$. Among the others, we mention a structure, defined by Eisenhart [8], that we have considered with particular emphasis in our previous papers [3–6,9]. In this case the ambient space is an enlarged configuration space-time $M \times \mathbb{R}^2$, with local coordinates $(q^0, q^1, \ldots, q^N, q^{N+1})$, where $(q^1, \ldots, q^N) \in M$, $q^0 \in \mathbb{R}$ is the time coordinate, and $q^{N+1} \in \mathbb{R}$ is a coordinate closely related to Hamilton action; Eisenhart defines a pseudo-Riemannian non-degenerate metric $g_E$ on $M \times \mathbb{R}^2$ as

$$ds^2_E = g_{\mu\nu} \, dq^\mu \otimes dq^\nu = a_{ij} \, dq^i \otimes dq^j - 2V(q) \, dq^0 \otimes dq^0 + dq^0 \otimes dq^{N+1} + dq^{N+1} \otimes dq^0 \; . \tag{6}$$

Natural motions are now given by the canonical projection $\pi$ of the geodesics of $(M \times \mathbb{R}^2, g_E)$ on configuration space-time: $\pi : M \times \mathbb{R}^2 \to M \times \mathbb{R}$. However, among all the geodesics of $g_E$ the natural motions belong to the subset of those geodesics along which the arclength is positive definite

$$ds^2 = g_{\mu\nu} dq^\mu dq^\nu = 2C^2 dt^2 \; . \tag{7}$$

The stability of a geodesic flow is studied by means of the Jacobi - Levi-Civita (JLC) equation for geodesic spread. In local coordinates the JLC equation reads as

$$\frac{\nabla^2 J^k}{ds^2} + R^k_{ijr} \frac{dq^i}{ds} J^j \frac{dq^r}{ds} = 0 \; , \tag{8}$$

where $R^k_{ijr}$ are the components of the Riemann-Christoffel curvature tensor. In previous papers we have investigated the relationship between geometry and chaos mainly using the Eisenhart metric described above. The JLC equation has been used in its exact form with Jacobi metric, only in the case of two degrees of freedom systems [2,1]: a perfect agreement between the description of instability provided by JLC equation and the description of instability provided by more conventional methods (Lyapunov exponents, Poincaré surfaces of section) has been found. Let us now extend our investigation to arbitrary $N$. To this purpose we use a natural chart (in previous works we adopted parallely trasported frames). Let us begin by computing the left hand side of Eq.(8). From $(\nabla J^k/ds) = dJ^k/ds + \Gamma^k_{ij} (dq^i/ds) J^j$ we have

$$\frac{\nabla^2}{ds^2} J^k = \frac{d}{ds}\left(\frac{dJ^k}{ds} + \Gamma^k_{ij}\frac{dq^i}{ds}J^j\right) + \Gamma^k_{rt}\frac{dq^r}{ds}\left(\frac{dJ^t}{ds} + \Gamma^t_{ij}\frac{dq^i}{ds}J^j\right) \tag{9}$$

trivial algebra and the use of Eq.(4) lead to

$$\frac{\nabla^2}{ds^2} J^k = \frac{d^2 J^k}{ds^2} + 2\Gamma^k_{ij}\frac{dq^i}{ds}\frac{dJ^j}{ds} + \left(\partial_r \Gamma^k_{ij} + \Gamma^k_{rt}\Gamma^t_{ij} - \Gamma^k_{tj}\Gamma^t_{ri}\right)\frac{dq^r}{ds}\frac{dq^i}{ds}J^j \tag{10}$$

where $\partial_i \equiv \partial/\partial q^i$. Then, we use the expression for the components of the Riemann-Christoffel tensor to obtain

$$R^k_{ijr}\frac{dq^i}{ds}J^j\frac{dq^r}{ds} = \left(\Gamma^t_{ri}\Gamma^k_{jt} - \Gamma^t_{ji}\Gamma^k_{rt} + \partial_j \Gamma^k_{ri} - \partial_r \Gamma^k_{ji}\right)\frac{dq^r}{ds}J^j\frac{dq^i}{ds} \tag{11}$$

and by substituting Eqs.(10) and (11) into Eq.(8) we finally get

$$\frac{d^2 J^k}{ds^2} + 2\Gamma^k_{ij}\frac{dq^i}{ds}\frac{dJ^j}{ds} + \left(\frac{\partial \Gamma^k_{ri}}{\partial q^j}\right)\frac{dq^r}{ds}\frac{dq^i}{ds}J^j = 0 \tag{12}$$

which has general validity *independently* of the metric of the ambient manifold. Let us now derive its explicit form in the case of Jacobi metric. This metric is a conformal deformation of the pure kinetic energy metric, i.e. $(g_J)_{ij} = e^{-2f} a_{ij}$. As we are mainly interested in studying standard Hamiltonian systems, $a_{ij} = \delta_{ij}$ is assumed. For a conformal metric $(g_J)_{ij} = e^{-2f}\delta_{ij}$ one readily obtains the following expression for the Christoffel coefficients: $\Gamma^k_{ij} = -\delta^k_j f_{,i} - \delta^k_i f_{,j} + \delta_{ij} f^{,k}$, where $f_{,i} = \partial_i f \equiv \partial f/\partial q^i$. Hence Eq.(12) is transformed into

$$\frac{d^2 J^k}{ds^2} - 2\frac{df}{ds}\frac{dJ^k}{ds} - 2\frac{dq^k}{ds}\frac{d}{ds}(f_j J^j) + 2f_k \frac{dq^i}{ds}\delta_{ij}\frac{dJ^j}{ds} + f_{kj} J^j e^{2f} = 0 \ , \tag{13}$$

and, using the relation $ds = e^{-2f} dt$, we can express it in terms of the physical time $t$ instead of the proper time $s$:

$$\frac{d^2 J^k}{dt^2} + 2\left(f^{,k}\delta_{ij}\frac{dq^i}{dt} - f_{,j}\frac{dq^k}{dt}\right)\frac{dJ^j}{dt} + \left(f_{,kj} e^{-2f} - 2f_{,ji}\frac{dq^i}{dt}\frac{dq^k}{dt}\right) J^j = 0 \ , \tag{14}$$

where $f_{,ij} = \partial^2_{ij} f$. Finally, as the Jacobi metric corresponds to $f = \frac{1}{2}\ln[1/2(E-V)]$, it is

$$f_{,i} = \frac{\partial_i V}{2(E-V)} \tag{15}$$

$$f_{,ij} = \frac{\partial^2_{ij} V}{2(E-V)} + \frac{(\partial_i V)(\partial_j V)}{2(E-V)^2} \tag{16}$$

$$e^{-2f} = 2(E-V) \tag{17}$$

so that the final expression for the JLC equation for $(M_E, g_J)$ is

$$\frac{d^2 J^k}{dt^2} + \frac{1}{E-V}\left(\partial_k V \delta_{ij}\frac{dq^i}{dt} - \partial_j V \frac{dq^k}{dt}\right)\frac{dJ^j}{dt}$$
$$+ \frac{1}{E-V}\left[(E-V)\partial^2_{kj} V + (\partial_k V)(\partial_j V) - \left(\partial^2_{ij} V + \frac{(\partial_i V)(\partial_j V)}{E-V}\right)\frac{dq^i}{dt}\frac{dq^k}{dt}\right] J^j = 0 \ . \quad (18)$$

Let us now give the explicit form of Eq.(12) in the case of $(M \times \mathbb{R}^2, g_E)$, the enlarged configuration space-time equipped with Eisenhart metric. One easily finds [9] that only the following Christoffel coefficients do not vanish: $\Gamma^i_{00} = (\partial V/\partial q_i)$ and $\Gamma^{N+1}_{0i} = (-\partial V/\partial q^i)$, hence, using also $ds^2 = (dq^0)^2 = dt^2$ (as we can set $2C^2 = 1$), we get

$$\frac{dJ^k}{ds^2} + \frac{\partial^2 V}{\partial q^j \partial q_k} J^j = 0 \qquad (19)$$

for $k, j = 1, \ldots, N$. The two other components, $J^0$ and $J^{N+1}$, do not contribute to the norm of $J$ and do not enter the evolution equation (19), therefore they can be neglected [9].

It is a very interesting fact that the JLC equation (8) yields the usual tangent dynamics equation (19) when explicitly worked out for the Eisenhart metric on $M \times \mathbb{R}^2$. On one hand we can expect that at least *qualitatively* Eq.(18) will give similar results to those obtained with equation (19), i.e. the usual Lyapunov exponents. On the other hand, the two equations (18) and (19) are so different that it is unclear whether a *quantitative* agreement too has to be expected. Geodesics of $(M \times \mathbb{R}^2, g_E)$ project themselves onto geodesics of $(M_E, g_J)$: for this reason unstable (stable) geodesics of $(M \times \mathbb{R}^2, g_E)$ must correspond to unstable (stable) geodesics of $(M_E, g_J)$. However, no theoretical result guarantees that the average growth-rates of the solutions of Eqs. (18) and (19) must coincide. We have addressed this point by numerically computing the average growth-rates of the solutions of Eqs. (19) and (18) – let us denote them by $\lambda_1$ and $\lambda_1^{JLC}$ respectively – for a given Hamiltonian flow with a large number of degrees of freedom; $\lambda_1$ is the conventional largest Lyapunov exponent.

Numerical computations have been performed for a flow described by the Hamiltonian

$$\mathcal{H}(p,q) = \sum_{i=1}^{N}\frac{1}{2}p_i^2 + \sum_{i=1}^{N}\left[\frac{1}{2}(q_{i+1}-q_i)^2 + \frac{\mu}{4}(q_{i+1}-q_i)^4\right] \ . \qquad (20)$$

This is the well known Fermi-Pasta-Ulam $\beta$-model [10], a paradigmatic model of non linear classical many-body systems extensively studied over the last decades and at the origin of remarkable developments in nonlinear dynamics (for instance, the transition between weak and strong chaos has been first discovered in this model [11,12]).

The numerical integration of the equations of motion (5) derived from the Hamiltonian (20) has been performed by means of a third order bilateral symplectic algorithm [13], and the integration of the two stability equations (19) and (18) has been done by means of the same bilateral algorithm and of a fourth-order Runge-Kutta scheme respectively. Both $\lambda_1(\epsilon)$ and $\lambda_1^{JLC}(\epsilon)$ have been obtained by means of a standard algorithm [14], i.e. computing

$$\lambda_1(t_\mathcal{N}) = \frac{1}{\mathcal{N}\Delta t} \sum_{n=1}^{\mathcal{N}} \ln \left( \frac{\|J(t_n)\|^2 + \|\dot{J}(t_n)\|^2}{\|J(t_{n-1})\|^2 + \|\dot{J}(t_{n-1})\|^2} \right) \qquad (21)$$

where $t_n = n\Delta t$, $\Delta t$ is some time interval, $t_\mathcal{N}$ is the final time such that $\lambda_1$ has attained a good "asymptotic" value.

In Fig.1 the values of $\lambda_1^{JLC}(\epsilon)$ are compared to the values of $\lambda_1(\epsilon)$ and to an analytically predicted curve for $\lambda_1(\epsilon)$ (see ref. [4]); $\epsilon = E/N$ is the energy density. As the numerical effort to integrate Eq.(18) is heavier than that required to integrate Eq.(19), we computed $\lambda_1$ for $N = 256$ and $N = 2000$ coupled oscillators, whereas we computed $\lambda_1^{JLC}$ for $N = 128$ and $N = 256$; at $N = 256$ we have only two points that have been computed just as a stability check. The excellent agreement between the outcomes of the two stability equations is well evident.

In Fig.2 the relaxation patterns of $\lambda_1^{JLC}(t)$ and of $\lambda_1(t)$ are also displayed. These are very similar at high energy density, whereas they show some separation at low energy density: the final values are nevertheless always in very good agreement. These results mean that equations (18) and (19) are not – loosely speaking – the "same" equation written in two different forms. As a matter of fact, Eq.(19) is contained in Eq.(18) so that one could think that in some non-trivial way the extra terms cancel out. This is not the case. There are two distinct equations to describe the same phenomenon. They are *equivalent* for what concerns the computation of the average instability growth rates of Hamiltonian flows, but they can be *not equivalent* for the further development of the theoretical approach where the average curvature properties of the "mechanical" manifolds are linked to the average chaoticity of the dynamics through an effective stability equation independent of the dynamics itself [4]. In fact Eq.(18) is valid on $(M_E, g_J)$, a manifold which has better mathematical properties with respect to $(M \times \mathbb{R}^2, g_E)$: $(M_E, g_J)$ is a proper Riemannian manifold, it is compact, all of its geodesics are in one-to-one correspondence with mechanical trajectories, its scalar curvature does not identically vanish as is the case of $(M \times \mathbb{R}^2, g_E)$, it can be naturally lifted to the tangent bundle where the associated geodesic flow on the submanifolds of constant energy coincides with the phase space trajectories.

In conclusion, we have seen that the results found for the $N = 2$ case [1,2] generalize to arbitrary $N$, hence the phenomenological information given by Lyapunov exponents can be retrieved on the manifold $(M_E, g_J)$ at arbitrary dimension by means of the JLC equation for geodesic spread.

# REFERENCES

†     E-mail: mcerruti@arcetri.astro.it

∗     Also at INFN, Sezione di Firenze, Italy. E-mail: franzosi@fi.infn.it

‡     Also at INFM, unità di Firenze, and INFN Sezione di Firenze, Italy. E-mail: pettini@arcetri.astro.it

## FIGURES

FIG. 1. $\lambda_1^{JLC}(\epsilon)$ computed at $N = 128$ is represented by full circles and computed at $N = 256$ by full triangles. The largest Lyapunov exponent $\lambda_1(\epsilon)$ is represented by open circles ($N = 256$) and open squares ($N = 2000$). The solid line is the analytic prediction for $\lambda_1(\epsilon)$ given in ref. [4].

FIG. 2. The relaxation patterns $\lambda_1^{JLC}(t)$ and $\lambda_1(t)$ are compared at different values of the energy density. Full symbols denote $\lambda_1^{JLC}(t)$ and open ones denote $\lambda_1(t)$. From top to bottom $\varepsilon = 392$, $\varepsilon = 1$, $\varepsilon = 0.075$.

M.Cerruti-Sola,R.Franzosi,M.Pettini
'Lyapunov exponents from geodesic...'
Fig. 1

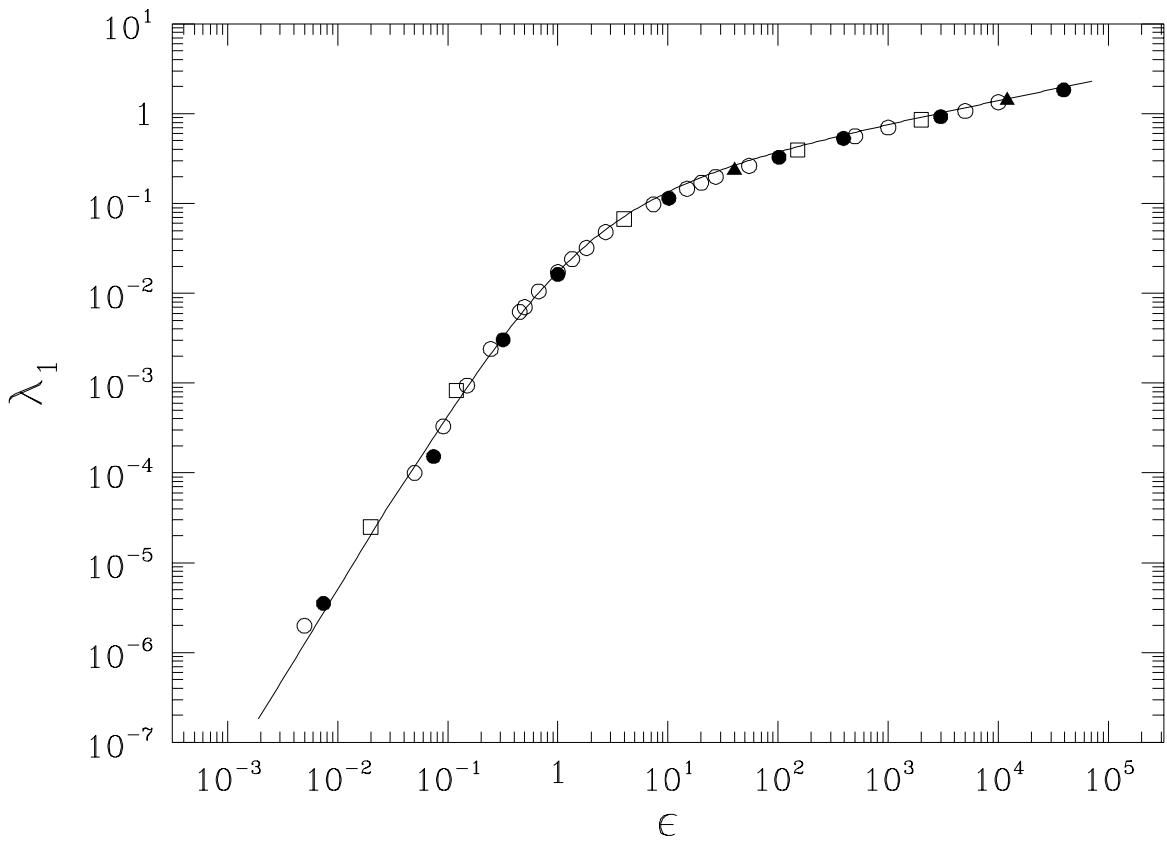

M.Cerruti-Sola, R.Franzosi, M.Pettini
'Lyapunov exponents from geodesic...'
   Fig. 2

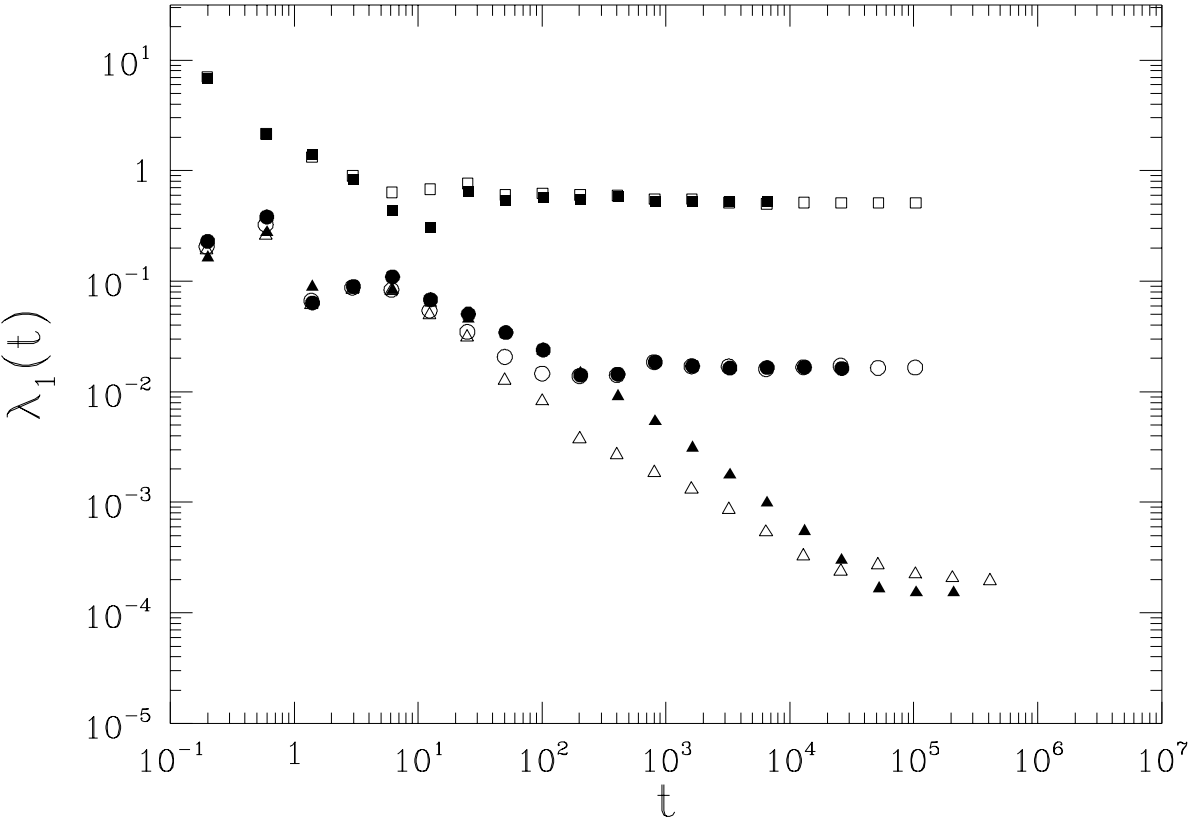